\documentclass[11pt,preprint]{aastex}

\newcommand{\msun}{M$_{\sun}$}
\newcommand{\uv}{{\it u,v}\ }
\newcommand{\lsun}{L$_{\sun}$}

\shorttitle{Sub-Arcsecond Modeling of YSOs}
\shortauthors{Looney, Mundy, \& Welch}

\begin{document}

\title{Envelope Emission in Young Stellar Systems:\\
A Sub-Arcsecond Survey of Circumstellar Structure}
\author{Leslie W. Looney\altaffilmark{1}}
\affil{Department of Astronomy, University of Illinois, Urbana-Champaign} 
\author{Lee G. Mundy}
\affil{Department of Astronomy, University of Maryland, College Park}

\and

\author{W. J. Welch}
\affil{Radio Astronomy Laboratory, University of California, Berkeley}

\altaffiltext{1}{Email: lwl@uiuc.edu}

\begin{abstract}
We present modeling results for six of the eleven deeply embedded systems from
our sub-arcsecond $\lambda$~=~2.7~mm continuum interferometric survey.
The modeling, performed in the \uv plane, 
assumes dust properties, allows for a power-law density profile, 
uses a self-consistent, luminosity conserving temperature profile, 
and has an embedded point source to represent a circumstellar disk.
Even though we have the highest spatial resolution to date
at these wavelengths, only the highest signal-to-noise systems can
adequately constrain the simple self-similar collapse models.
Of the six sources modeled, all six were fit with a density power-law
index of 2.0; however, in half of the systems, those with the highest
signal-to-noise, a density power-law index of 1.5 can be rejected at the
95\% confidence level.
Further, we modeled the systems using the pure Larson-Penston (LP) and Shu solutions
with only age and sound speed as parameters.
Overall, the LP solution provides a better fit to the data, both in likelihood and
providing the observed luminosity, but the
age of the systems required by the fits are surprising low
(1000-2000 yrs).
We suggest that either there is some overall time scaling of the 
self-similar solutions that invalidate the age estimates, or more likely
we are at the limit of the usefulness of these models.
With our observations we have begun to reach the stage where models
need to incorporate more of the fundamental physics of the collapse
process, probably including magnetic fields and/or turbulence.
In addition to constraining collapse solutions, our modeling allows the
separation of large-scale emission from compact emission, enabling the
probing of the circumstellar disk component embedded within the protostellar
envelope.
Typically 85\% or more of the total emission is from the extended circumstellar
envelope component. 
Using HL Tauri as a standard candle, the range of circumstellar disk masses
allowed in our models is 0.0 to 0.12 \msun;
our Class 0 systems do not have disks that are significantly more massive than 
those in Class I/II systems.
This implies that the disk in Class 0 systems must
quickly and efficiently process $\sim$ 1 \msun\ of material from
the envelope onto the protostar.

\end{abstract}


\section{Introduction} \label{intro}

Stars form deep within dense molecular cores which
provide the raw material for star formation. Their structure and
kinematics determine how the formation proceeds and the
eventual outcome: single star, binary, or multiple
system. For isolated low mass stars, we have a good overall
picture of the process and well developed models for an idealized
collapse to form a single star system.  At a broad-brush level,
these idealized collapse models are able to provide
a good understanding of the process, despite shortcomings
with respect to stellar multiplicity and core geometry:
dense cores are roughly spherical and stellar multiplicity typically
occurs on scales much smaller than scales characteristic of cores.
This paper utilizes high resolution observations of the dust continuum
emission from cores around young stars to study the morphology of the
inner cores, or {\it envelopes}, around young systems. The primary aim is
to compare the inferred density structure to that expected from 
collapse models.

The solution to the collapse of an isothermal sphere can
be thought of as a continuum of 
solutions \citep[but also see Hunter 1986]{whit}
bracketed by the two standard solutions:
the Larson-Penston solution, or LP solution, put forth by
\cite{larson69} and \cite{penston} and expanded by \cite{hunter} 
and the Shu solution proposed by \cite{shu77}.
The LP solution is valid for the isothermal spherical
collapse of an initially uniform medium. The core density profile
is characterized by a uniform density
central region surrounded by a $\rho \propto r^{-2}$ envelope.
As collapse progresses, the uniform density region shrinks until the
entire sphere follows $\rho \propto r^{-2}$; continued collapse
sets up an $r^{-3/2}$ central density profile around the
central forming star \citep{hunter}. In the other line of heritage,
\cite{shu77} defined a class of self-similar collapse solutions 
based on an initial singular isothermal spherical density distribution, 
$\rho = \frac{a^{2}}{2\pi G}r^{-2}$,  where
a is the local sound speed in the gas
and G is the gravitational constant.
The sphere collapses most rapidly at the center, leading to
a collapse wave which moves outward in radius at the local sound speed. 
This is referred to as an ``inside-out'' collapse; the density profile inside
the wave approaches a free-fall density profile, $\rho \propto r^{-3/2}$;
outside the collapse wave, the profile remains $\rho \propto r^{-2}$.
Both solutions have density profiles of $\rho \propto r^{-2}$
in the outer core and both tend toward
$\rho \propto r^{-3/2}$ in the centers some time after the 
star begins forming. 

The biggest differences between the two solutions
involve the mass infall rate, infall velocity structure, 
and the density structure evolution. 
The Shu solution has a
constant mass infall rate at all times; the mass in the central object is 
nearly equal
to the mass within the collapse wave radius at any time; and material
outside the collapse wave is nearly static because the collapse starts from
a static isothermal sphere.
The LP solution has collapse motions wherein all radii acquire a velocity of
3.3 times the sound speed,
a mass infall rate that peaks at the beginning then asymptotically approaches 
the Shu value over time, and a maximum infall velocity that is not in the center, as 
in the Shu solution, but at a finite radius. 

\cite{me} (hereafter LMW) presented
images of 24 young stellar sources with sensitivity to spatial scales
from 0$\farcs$5 to 50$\arcsec$ in the $\lambda$~=~2.7~mm continuum.
A discussion of the data acquisition and images can be found in that
paper.
In this paper, we present modeling of the envelope emission 
for many of the embedded objects in LMW.
Our primary goals in the modeling are to address three issues:
(1) What constraints, if any, can we place on the power-law of the density
and on the two collapse solutions?  
Does the power-law index resemble the isothermal sphere ($\rho \propto r^{-2}$)
or the free-fall profile ($\rho \propto r^{-3/2}$)?  Or something else
entirely?  
For example, \cite{ward} found that starless clouds tended to
have flat-topped density profiles, more like Bonner-Ebert spheres.
(2) What constraints can we place on the circumstellar disks in these
systems?
With the highest resolution to date at these wavelengths, we will
be able to place limits on the size of the embedded disk.
Can the data be fit with or without a central circumstellar disk?
(3) What constraints can we place on the inner and outer
radii of the envelopes?

\section{The Observational \uv Data} \label{obs}

To compare the observations and models, we analyze the data in
\uv space where the data are not effected by the CLEAN algorithm
or \uv sampling.
The interferometric data are binned in \uv annuli around the 
source location from LMW and averaged vectorially.
The resulting \uv amplitudes for the bins are 
shown in Fig. \ref{uvdata}
with $\log$(\uv distance) versus $\log$(amplitude) plots.
The displayed error bars are statistical error bars
based on the standard deviation of the mean of the data points in the bin
with a minimum of 10\%, reflecting the uncertainty in the
overall calibration. 
In the cases where the binary systems were separated by more 
than 10$\arcsec$, the companion sources were subtracted out of the
\uv data using the large scale images of LMW.
The new \uv data were remapped to confirm that the large scale emission from
the companion sources was not detected.
Although there may exist some residual of small-scale emission from
the companion envelope in the \uv data, vector averaging in \uv annuli 
will minimize its contribution.

For most sources, the circular symmetry of the observed emission and
the lack of significant internal structure
(see images of the Class 0 sources in LMW)
justifies the simplicity of a spherical model.
The spherical envelope can be characterized by two radial distributions:
$\rho(r)$ and $T(r)$.  These quantities are typically and reasonably assumed to be
power-laws in radius \citep{adams88,beck90,adams90,keene90,beck91,sue93}
$\rho(r) = \rho_{o}(r/r_{o})^{-p}$ and $T(r) = T_{o}(r/r_{o})^{-q}$.

What shape do we expect for the classic envelope power-law emission
model in the \uv plane?
One can get a sense from simple consideration of the 
interdependence of the radial 
brightness distribution in the image and {\it u,v} planes.
A power-law radial dependence of the emission in the 
image plane, $B(r)$,
transforms to a power-law dependence of the
visibility amplitudes on \uv distance, $V(s)$.
The index of the power-law in \uv distance is
related to the index of the power-law in radial distance.
Specifically, as long as the emission is optically thin and in the Rayleigh-Jeans
limit, and the envelope infinite in size, 
$$ B(r) \propto r^{-(p+q) + 1}\ \ \rightarrow\ \
V(s) \propto s^{(p+q-3)}$$
where $s$ is \uv distance \citep[cf.][]{diss}.
Note that the slope is only sensitive to the sum of the
temperature and density power-law indices.
In the simplest terms, the large \uv spacings convey information
about the inner part of the envelope and the short \uv
spacing reflect the structure of the outer portion of the envelope.
If the envelope is truncated at a finite radius,
the cutoff in radius is equivalent to the convolution
of a modified first-order Bessel function with the power-law
in the \uv plane,
causing a flattening of the slope at short \uv spacings and
a ringing effect in the \uv distance versus amplitude plot.
This ringing arises from the sharp edged
envelope which is probably not physical, but the short spacing
turnover is robust to the slope of the cutoff.

The bottom right panel of Fig. \ref{uvdata}, shows
two example simple power-law models-- a finite envelope and
compact circumstellar disk at the distance of Taurus.
The envelope has an outer radius of 5000 AU, a density profile of
$\rho \propto r^{-2}$, and a temperature profile of $T \propto r^{-0.4}$.
The disk has an outer radius of 100 AU, a surface density profile
of $\Sigma \propto r^{-1.5}$,
and a temperature profile of $T \propto r^{-0.5}$.
The disk model shows 
that the flux remains
constant with increasing \uv distance
until the \uv distance is large enough to resolve 
the radial structure in the source.
For the envelope, its finite size also creates
a rounding off of the curve at very small \uv spacings,
as the source becomes unresolved.
When this happens the flux transitions to follow
the power law with index $p+q-3$.
However, since the envelope is larger than a disk, the related power-law
of the sum of the temperature and density is more quickly seen
at shorter \uv spacings.

Comparing the panels in Fig. \ref{uvdata}, it 
is immediately obvious that the data and models are not well 
represented by a single power-law extending to all radii.
The most significant deviation can be seen
at short \uv distances where the flux levels off in most sources,
indicating finite sized envelopes.
In two cases, IRAS 4 A and IRAS 16293-2422, the shortest
\uv data, corresponding to the largest spatial scales,
rises abruptly.
For these two objects, it is most likely that there is 
some contamination from the very large scale cloud
within which the sources are embedded.
In addition, IRAS 2 B and IRAS 4 C do
not follow the expected profile of an envelope dominated source
and may be either very compact envelopes or circumstellar disks.
This was also suggested by their near unity 5 k$\lambda$/50 k$\lambda$ ratio
in LMW.
As the very smallest spatial scales (the
largest \uv distances) are probed, 
there are two separate profiles
that stand out.
Some sources (e.g. IRAS 4A or L1448 B)
are consistently dropping in amplitude with
increasing \uv distance, suggesting that the 
power-law is being traced down to very small scales.
Others (e.g. SVS 13 A or IRAS 2A) flatten toward 
a constant flux level at very small scales, perhaps
suggesting that there is an unresolved point source
at the center of the envelope, arguably
a circumstellar disk.

\section{Sources and \uv Data}

\subsection{L1448 IRS3}

L1448 IRS3 \citep{bach86} is comprised of three
distinct sources in the $\lambda$~=~2.7~mm continuum
\citep[Fig. 7a LMW]{sue93,sue97}.
Sources A
and B, a proto-binary system (separation $\sim$ 7$\arcsec$), share
a common large-scale envelope.  Only source B provides sufficient
signal-to-noise in the \uv plane to be modeled effectively.  
However,
the data for source C in Fig. \ref{uvdata} drops sharply in amplitude
with \uv distance implying that the source is extended and dominated by
the envelope component.

For the modeling, we assume that the circumbinary envelope is centered on
source B.  The total luminosity of sources A and B is estimated to be 8
\lsun \citep{bar98}; we assume that the luminosity
of source B is 6.8 \lsun, based on the ratio of fluxes of sources A and
B at $\lambda$~=~2.7~mm (LMW).
Fig. \ref{uvdata} displays the \uv data for L1448 IRS3 B, and the
data are smooth with a well defined slope of
-0.3 within the inner 40 k$\lambda$
(corresponding to $p+q~\sim$~2.7, see \S \ref{obs}).
At 40 k$\lambda$, the
curve transitions into a steeper slope of -1.1
(corresponding to $p+q~\sim$~1.9, see \S \ref{obs}).

\subsection{NGC 1333 IRAS 2}

The young, binary system NGC 1333 IRAS2 
\citep[LMW]{jenn87,blake97}
has a separation of $\sim$ 31$\arcsec$.
In our high resolution image (Figure 8, LMW),
the envelope of source A appears to be completely resolved out, and the
remaining emission is consistent with a point source.
Source B (Fig. 9, LMW) appears to be slightly resolved,
and we concluded from Fig. 17 in LMW that source B is a very compact
source.
Overall, there may be an extended envelope that surrounds both of the sources,
but our data can not place useful limits on this structure.
The luminosity for sources A and B is estimated to be 40 \lsun
\citep{jenn87}.
The luminosity
for source A is assumed to be 30 \lsun, based on the ratio of fluxes
of sources A and B at $\lambda$~=~2.7~mm (LMW).

Fig. \ref{uvdata} displays the \uv data for NGC 1333 IRAS 2 A,
after subtraction of source B from the \uv data.
The curve has a well defined slope of
-0.4 within the inner 12 k$\lambda$ (corresponding to
$p+q~\sim$~2.6, see \S \ref{obs}).
Unlike L1448 IRS3 B, the visibility at larger \uv distance
flattens to a constant flux, rather than sharply descending.
This indicates that the model point source component will play
an important role.

As suggested in Fig. 17 of LMW, IRAS2 B is a compact source;
the amplitude does not significantly drop off until around 100 k$\lambda$.
It seems likely that this source has either a small, compact envelope
or has a circumstellar disk with very little envelope emission.
The slight dip in the visibility at 20 k$\lambda$ may be an artifact from the
subtraction of source A from the \uv data.

\subsection{SVS 13}

SVS 13 \citep{strom76} has at least four young stellar systems 
(LMW).
Besides the large-scale envelope that may surround
all of the sources, A and B are embedded in individual
envelopes, and, in the case of source A, the separate
envelope is probably a circumbinary envelope that enshrouds both A1 and A2
(separation of $\sim$5$\arcsec$).
In the $\lambda$~=~2.7~mm high resolution images (Figs. 10 \& 11, LMW),
SVS 13 A1 and B are both amorphous structures.
The total luminosity for sources A, B, and C is 45 \lsun\
\citep{jenn87}; we assume that the luminosity
for source A and B are 19 and 22 \lsun, respectively,
based on the ratio of fluxes
of sources A and B at $\lambda$~=~2.7~mm (LMW).

For source A, Fig. \ref{uvdata} shows the \uv data flatten to a 
constant in the outer
\uv spacings.  The curve has a well defined slope of -0.2 within
the inner 8 k$\lambda$ (corresponding to $p+q~\sim$~2.8, see \S \ref{obs}).
At 8 k$\lambda$, there is an inflection point, and the curve
transitions to a steeper slope of -0.8 (corresponding to 
$p+q~\sim$~2.2, see \S \ref{obs}).

In Fig. \ref{uvdata}, the \uv data are shown for NGC 1333 SVS 13 B.
There is a slight ``bump'' in the curve at 80 k$\lambda$, which corresponds
to a fringe of around 3$\arcsec$.
This excess in amplitude may be a beating effect from incomplete
subtraction of one of the other sources,
or the excess may be from a non-spherical symmetric component of source B on
a size scale of $\sim 3\arcsec$.
In either case, the 80 k$\lambda$ \uv point cannot be fit with a
simple-model, increasing the reduced $\chi^2$ and biasing the amplitude upward
in the outer parts of the \uv plane.
The slope of the \uv data from 5~k$\lambda$ to 35~k$\lambda$ is -0.6
(corresponding to $p+q~\sim$~2.4, see \S \ref{obs}).

\subsection{NGC 1333 IRAS4}

NGC 1333 IRAS4 is comprised of
at least four distinct young stellar objects \citep[LMW]{sand91,lay95}.
Source A, the northern source, is
a 1$\farcs$7 binary system that shares a common circumbinary envelope
\citep[LMW]{lay95}.   
Source C is a compact source that has a brightness distribution that
more resembles an optical/IR source than its IRAS 4 companions
(LMW).
The total luminosity for sources A and B is assumed to be 21 \lsun\
\citep{jenn87}, and the luminosity 
for source A and B is assumed to be 16 and 5 \lsun, respectively,
based on the ratio of fluxes
of sources A and B at $\lambda$~=~2.7~mm (LMW). 

\begin{figure}
\includegraphics{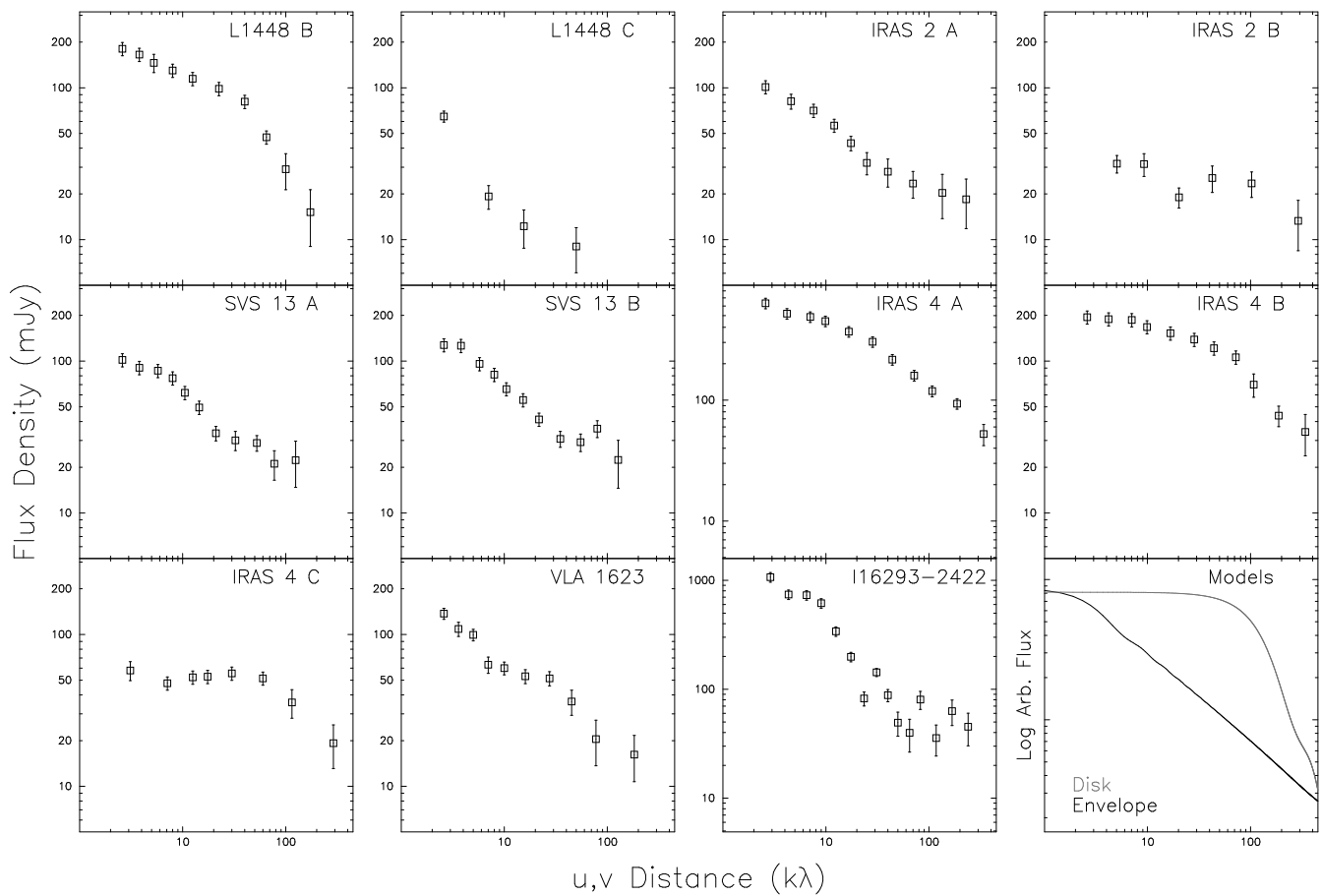}
\vspace{8cm}
\caption{
The \uv flux density data for each source, after subtraction of nearby companions,
averaged in annuli around the center position.  The panels are labeled with
the source name.  
The open squares are the data; the vertical lines are 1-sigma error
bars.  
The panel in the lower right shows the \uv dependence for
a power law envelope (solid) and disk (dashed); see \S \ref{obs} for more details.
}
\label{uvdata}
\end{figure}

In Fig. \ref{uvdata}, the \uv data are shown for NGC 1333 IRAS 4 A1---
the brighter source at $\lambda$=2.7mm in the 1$\farcs$7 binary.
We did not subtract out the binary companion source A2; running test
models with a second fixed point source at the location of source
A2 does not significantly alter the fits.
The shortest \uv spacing data point in Fig. \ref{uvdata} is
excessively high.
The best explanation is that the data are beginning to pick-up very
large-scale structure, perhaps from the cloud.
Since we are not including such a large-scale envelope in this modeling,
we only use the shortest \uv spacing point as a maximum constraint in the
models.
In that case, the data has a well defined slope of -0.3 in the 3 k$\lambda$
to 35 k$\lambda$ range (corresponding to
$p+q~\sim$~2.3, see \S \ref{obs}) and a slope of -0.7
in the 22 k$\lambda$ to 125 k$\lambda$ range (corresponding to
$p+q~\sim$~1.9, see \S \ref{obs}).

In Fig. \ref{uvdata}, the \uv data are shown for NGC 1333 IRAS 4 B.
The data has a well defined slope of -0.2 in the 5 k$\lambda$
to 90 k$\lambda$ range (corresponding to
$p+q~\sim$~2.8, see \S \ref{obs}).
Unlike the data from IRAS 4 A, the slope in the outer \uv spacings
is particularly steep.
In Fig. \ref{uvdata}, the \uv data are shown for NGC 1333 IRAS 4 C, after
subtraction of sources A and B from the \uv data.
As stated in LMW, the source is mostly compact with nearly constant
\uv amplitude from 2 k$\lambda$ to 80 k$\lambda$.
As shown in Fig. \ref{uvdata}, IRAS 4 C becomes resolved for
\uv distances longer than
80 k$\lambda$, but there are not enough data in this region to
constrain the density power-law.

\subsection{VLA 1623}

VLA 1623 is the prototype Class 0 source \citep{andre93}.
However, our high resolution $\lambda$~=~2.7~mm observations (Figure 15, LMW)
suggest that the system is actually a binary system with a separation of
1$\farcs$1.
The \uv data were binned around the center of the system, between the
two point sources.
The first three \uv spacing data points are amplitude biased
by large scale emission from the nearby regions of SMM1 and SMM2
\citep{ward89,andre93}.
From 7 k$\lambda$ to 25 k$\lambda$, the circumbinary envelope structure
dominates the \uv data, with a shallow slope of -0.1
(corresponding to $p+q~\sim$~2.9, see \S \ref{obs}).
Although, the exact index could vary significantly based upon
embedded point source(s), inner radius, and outer radius.
At \uv spacings greater than 30 k$\lambda$, the two circumstellar regions
beat against each other, making it difficult to estimate the
density power-laws.

We attempted to fit the \uv data with two point sources of
flux equal to the peak flux in Figure 15 of LMW.
However, this model overestimated the flux in the outer \uv spacings.
A correct model requires two circumstellar disks (or point sources)
embedded within
two circumstellar envelopes, and perhaps a larger circumbinary envelope.
Since this type of model has too many free parameters to be constrained
by the current data, VLA 1623 was not modeled further.

\subsection{IRAS 16293-2422}

IRAS 16293-2422 is a deeply embedded binary with two molecular outflows
\citep{walker86,woot89,mundy92}.
In high resolution observations at $\lambda$~=~2~cm, source A has two peaks,
A1 and A2 \citep{woot89}.
In the $\lambda$~=~2.7~mm high resolution image (Figure 16, LMW),
source A and B are clearly detected.
Source A, the most extended object in the survey, appears slightly elongated along the
position angle of the $\lambda$~=~2~cm sources.
The two sources, which may have individual circumstellar envelopes and disks,
are embedded within a circumbinary envelope.

The \uv data
were binned around source B, since source A was known to be
very extended.
Like NGC 1333 IRAS 4 A, the first \uv data point is biased high,
due to a large scale structure.
There is a very steep slope of -1.7 from 8 k$\lambda$ to 20 k$\lambda$, which
indicates a small $p+q$ value of 1.3.
For the \uv values greater than 20 k$\lambda$, the beating effect of the
two sources dominates the curve.
We attempted to fit the \uv data
with two point sources of
flux equal to the peak flux in (Figure 16, LMW).
As in the case of VLA 1623, this model overestimated the
flux in the outer \uv spacings.
A more complicated model is required, but these data would not
constrain the model; thus, IRAS 16293-242 was not modeled further.

\section{Modeling of the Emission} \label{kappa} \label{wc}

The simple power-law discussed in \S \ref{obs} provides
the first insights into the behavior of the brightness
distribution in the \uv plane. However,
detailed physical interpretation of the data requires
radiative transfer modeling.
Unfortunately, a number of assumptions must be made about the dust properties and source
morphology in order to simplify the
model and reduce the number of free parameters; they make the model
calculation easier but also reflect the limitations of the observational
data and of our knowledge of dust and source properties.
In the spirit of modest complexity, our basic model is
a spherical dust envelope with an embedded emitting point source, representing
an unresolved circumstellar disk.

\subsection{Dust Properties} \label{dustprop}
Dust emissivity is dependent on the
dust grain composition, size, and distribution
\citep[c.f.][]{krugel,pollack},
all of which can vary significantly with environment. 
Since grain emissivity affects grain thermal 
equilibrium and the observable emission, the determination 
of the dust temperature, column density, and optical depth
from the observed emission is explicitly dependent on 
knowledge of the underlying dust properties. 

For the broad dust description we adopt the model from  \cite{mark}
which uses a MRN \citep{mnr} dust grain distribution (``bare'' graphite
plus silicate grain distribution) and the \cite{draine} optical
constants to describe grain properties.
However, for this grain model the long wavelength dust emissivity
is assumed to follow a power-law ($\kappa_{\nu} \sim \lambda^{-\beta}$)
with $\beta$~=~2.
Studies have suggested that the dust emissivity in circumstellar
disks and dense cores has a $\lambda^{-1}$ dependence at submillimeter
wavelengths \citep{wein89,beck90,beck91} rather than the $\lambda^{-2}$
dependence found in calculations for bare grains  \cite[cf.][]{draine90}
and as observed in the large scale ISM \citep{reach}. 
Accordingly, we have modified the long
wavelength characteristics of our dust model ($\lambda~>~100~\mu$m) to
assume a $\lambda^{-1}$ emissivity. This was done by adopting a value of
$\kappa_{\nu}$(1~mm)~=~0.0033~cm$^{2}$ g$^{-1}$
and following that back to shorter wavelengths where
the $\lambda^{-1}$ curve intersected the bare grain emissivity curve.
This hybrid model preserves the optical and infrared properties of
the MRN dust grain model, while
forcing the long wavelength $\lambda^{-1}$ behavior and
$\kappa_{\nu}$(2.7~mm)~=~0.009~cm$^{2}$ g$^{-1}$.
Recent works looking at the $\kappa_\nu$(1.2~mm) appropriate for dense
molecular clouds and circumstellar disks yield reasonable consistent
values, $\kappa_\nu$(1.2~mm) = 0.004 cm$^{2}$ g$^{-1}$ with range of
a factor of 2 \citep{kramer}.
Our mass determinations, all based on $\lambda$~=~2.7~mm data,
depend roughly linearly on the assumed $\kappa$. 
We assume that dust properties are not a function
of radius in the envelope. 

\subsection{Temperature and Density Profiles}

As discussed in \S \ref{obs}, the young protostellar envelope
can be described by two radial power-law distributions for
density and temperature:
$\rho(r) = \rho_{o}(r/r_{o})^{-p}$ and $T(r) = T_{o}(r/r_{o})^{-q}$.
A power-law is a reasonable first-order approximation
for the density profile in many situations, as discussed in
\S \ref{intro}. Also, in a realistic system
the density distribution
has an inner and outer cutoff radius, $R_{in}$ and $R_{out}$. These
cutoffs are physically motivated by the presence of the circumstellar disk
which truncates the inner envelope and the outer
radius which is set by either a companion forming star or
merging with the extended cloud.

Although a power-law is a reasonable assumption for the density
profile, a radial power-law temperature profile is more problematic
for our application.
The temperature profile in an optically thin, dust envelope
heated by a central star will have a power-law
dependence on radius and stellar luminosity ($L_{*}$)
if dust opacity has a power-law dependence on frequency:
$T(r) = T_{o}(r/r_{o})^{-q}(L_{*}/L_{o})^{\frac{q}{2}},$
where $T_{o}$ is the dust temperature at the radius $r_{o}$ for
a stellar luminosity of $L_{o}$. 
The index $q$ depends on the dust opacity power-law
index, $q~=\frac{2}{4+\beta}$.
For values of $\beta$ from 0 to 2, the temperature
power-law index is between 0.33 and 0.5.

The use of simple power-laws in density and temperature allow for
a quick interpretation of an observed emission profile, as shown in 
\S \ref{obs}.
However, the temperature profile can diverge from a simple 
power-law as the envelope becomes optically thick at the primary
wavelengths of energy transport.
For a centrally peaked density profile,
the envelope can become optically thick over the inner 100's of AU,
resulting in a steeper inner temperature profile. To account for
the effects of this heating we have calculated 
self-consistent dust temperature profiles using the code of
\cite[WC hereafter]{mark}.

The WC code assumes a central heating source embedded within
a spherical dust envelope.  The central source is characterized by a
luminosity and an effective temperature.
For the effective temperature, we used 10,000 K, which is both
consistent with the effective temperature from
models of zero age low mass stars \citep{hejlesen,girardi} and
the temperature derived for T~Tauri stars to explain the veiling continuum
\citep{hartigan95}.
The dust envelope is specified by an outer radius, the power-law density
index, the density at the outer radius, and the destruction temperature
of the dust, which specifies the inner radius.
Given these parameters,
the WC code self-consistently calculates the dust temperature profile
by conserving luminosity at all radii.

\begin{figure}
\includegraphics{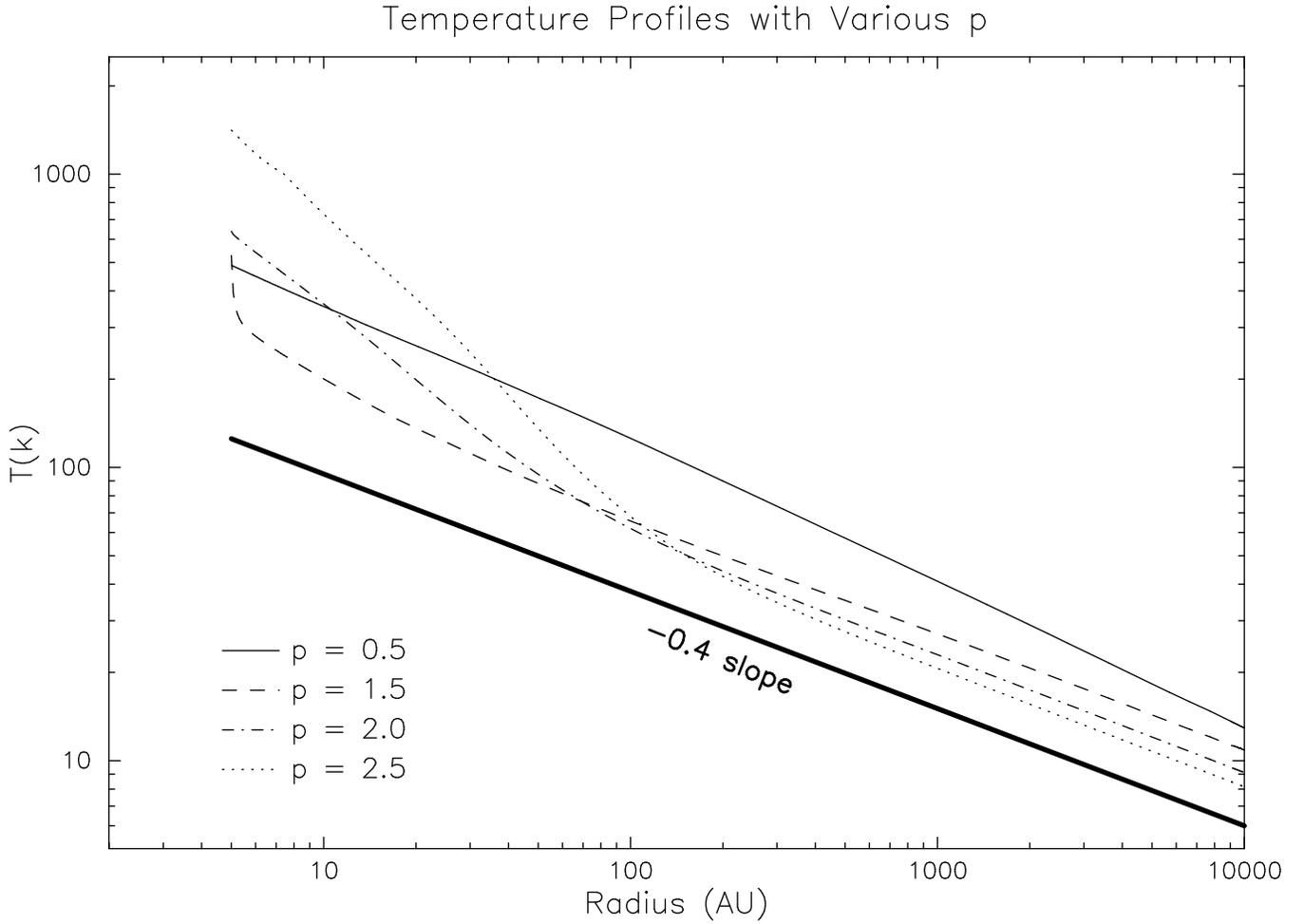}
\vspace{8cm}
\caption{
The dust temperature profile for a range in density power-law index (p).
The envelope has a mass of 1 \msun, a 10000 AU outer radius and a 
luminosity of 10 \lsun.
The dust emissivity is assumed to be a power law with $\beta~=~1$ 
for $\lambda > 100 \mu m$
and MRN dust for $\lambda < 100 \mu m$ (see \S \ref{dustprop}). 
The thick solid line indicates the slope of the emission expected for 
optically thin $\beta = 1$ dust.
The $p=0.5$ model (thin solid line) is nearly optically thin to the 
stellar radiation.
}
\label{diffp}
\end{figure}

\begin{figure}
\includegraphics{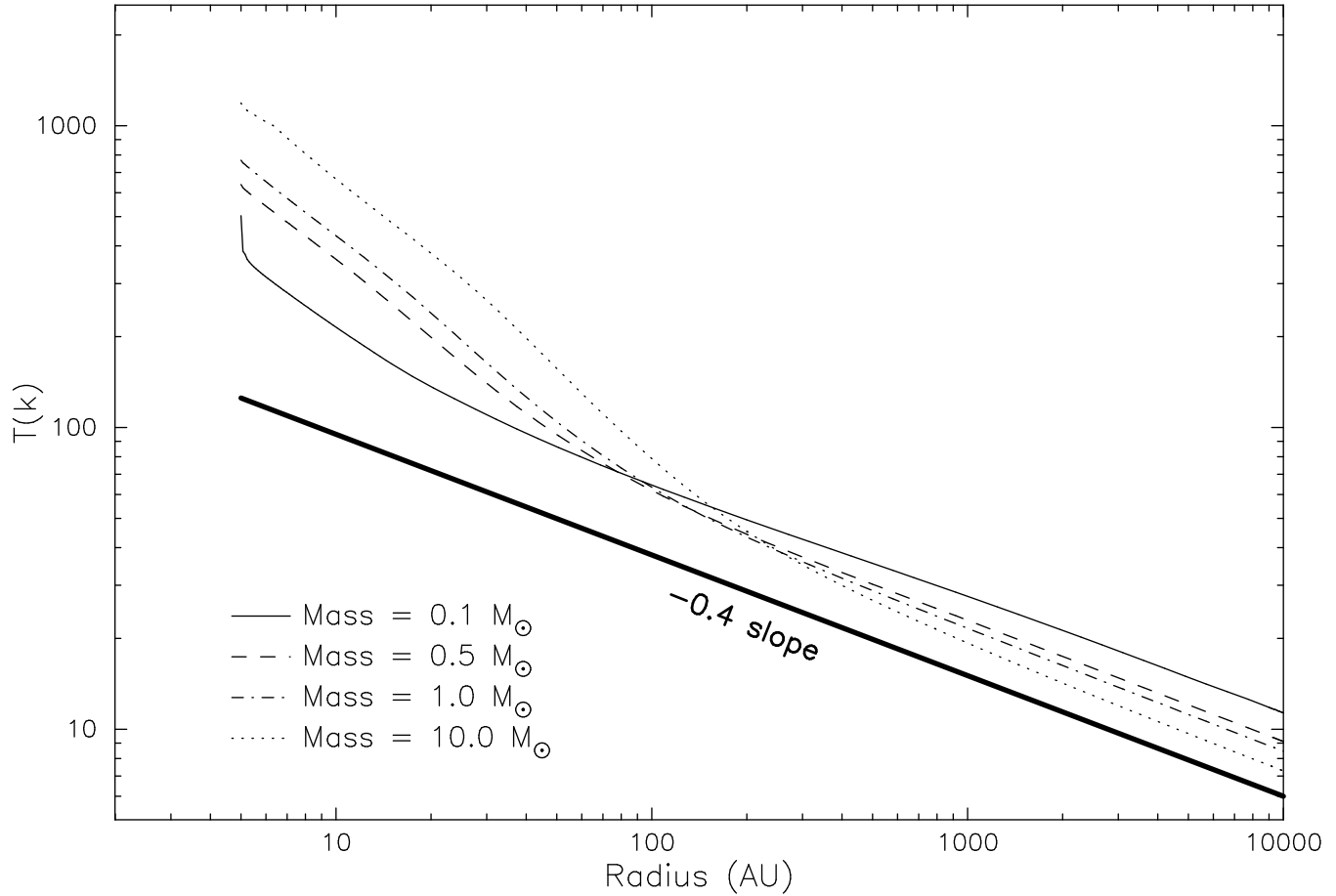}
\vspace{8cm}
\caption{
The dust temperature profile for a range of envelope masses with
a density power-law index of 2.0, a radius of 10000 AU, and a luminosity 
of 10 \lsun.
The dust properties are the same as for Figure \ref{diffp}. 
The thick solid line
indicates the slope of the emission expected for 
optically thin $\beta = 1$ dust.
}
\label{diffm}
\end{figure}

Figs. \ref{diffp} and \ref{diffm} display plots of the 
temperature profile for various values of the density power-law 
index p and envelope mass, respectively.
As can be seen in Fig. \ref{diffp}, the density power-law index
can have significant impact upon the temperature profile at radii 
less than 200 AU.
Indeed, the temperature radial distribution
can depart significantly from the optically-thin assumption of 
$T \propto r^{-0.4}$, the line at the bottom of the figure.
As the density power-law index increases, the inner opacity increases,
and the interior material can not radiate away its luminosity efficiently,
causing a steep increase in temperature.
At the outer radii, the emission escapes easily and
the temperature approaches the optically thin slope although
the absolute temperature can differ by as much as 50\%.
A similar effect occurs with increasing mass (Fig. \ref{diffm}).
As the mass of the envelope is increased the radius at which the
temperature approaches the optically thin limit increases.
The temperature at inner radii
can be as much as a factor of three times the optically
thin solution.

The WC code shows that the optically thin temperature
assumption is incorrect for envelope masses $>$~0.1~\msun\ and
density profiles steeper than p~=~1.
The largest changes occur at small radii,
$<$ 200 AU, (0$\farcs$6 at the distance of Perseus),
a region beginning to be probed by our observations.
Beyond 200 AU all of the temperature profiles tend
toward the optically thin $T \propto r^{-0.4}$ but with
a shift in the absolute temperatures.
Finally, increasing luminosity for fixed
envelope properties approximately scales the temperature profiles
by $(L/L_{\sun})^{0.25}$ as found by
\cite{dave95}.


\section{Modeling Details and Results}

For each model, the dust temperature distribution is derived
self-consistently with the luminosity;
the envelope emission is calculated as an image;
the central point source flux, attenuated by
the envelope, is added; and the model is multiplied by the BIMA
primary beam.
The resulting image is Fast Fourier Transformed 
and sampled with the same \uv spacings as the data.
The data and model are both vector-averaged
into \uv distance bins (annuli in the \uv plane) with the phase
center at the source position given in LMW.
A reduced $\chi^2$ of the fit is calculated using the
model, observed flux amplitudes, and the $\sigma$ of the
data in each radial bin.

The model has three fixed quantities (luminosity, dust emissivity, and
distance), as well as five degrees of freedom: power-law index of the density ($p$),
total mass of the envelope ($M$), inner cutoff radius ($R_{in}$),
outer cutoff radius ($R_{out}$), and point source flux ($S_{p}$).
For each model, the $\chi^2$ was minimized with respect to the envelope
mass to produce the best fit to the data.
We were only able to fully model six of the eleven
sources due to either a lack of signal-to-noise (L1448 C),
inability to constraint fit (IRAS 2B and IRAS 4C), or dominant
binarity (VLA 1623 and IRAS 16293-2422).

In this section, we will first examine the constraints
that the data can place on the power-law index of the density,
then, in more detail, the interesting $p~=~1.5$ and $p~=~2.0$
power-law indices, and finally, specific fitting of the 
Larson-Penston and Shu collapse solutions to the data.
In the latter case, the envelope parameters are uniquely
determined by the sound speed and the
time since the formation of a centralized source, t~=~0 in
the standard collapse scenario 
\citep[cf.][]{larson69,penston,shu77,hunter}.

\subsection{Results for Power Law Envelopes}

The first question to examine is how well does the data 
constrain the models parameters? 
The most interesting parameter is the density
power-law index since it is of central interest to collapse models.
To answer this, we explored parameter space
including: $p$ from 0.5 to 2.9 in steps of 0.2,
$R_{out}$ from 2000 AU to 10000 AU in steps of 1000 AU,
$R_{in}$ of 10 AU and 100 AU, and a central point source 
flux ($S_{ps}$) in 1$\sigma$ steps ($\sigma$~=~RMS noise 
in the last \uv distance amplitude bin) starting with no 
point source, and increasing to a maximum of the amplitude in 
the last \uv distance bin.

\begin{figure}
\includegraphics{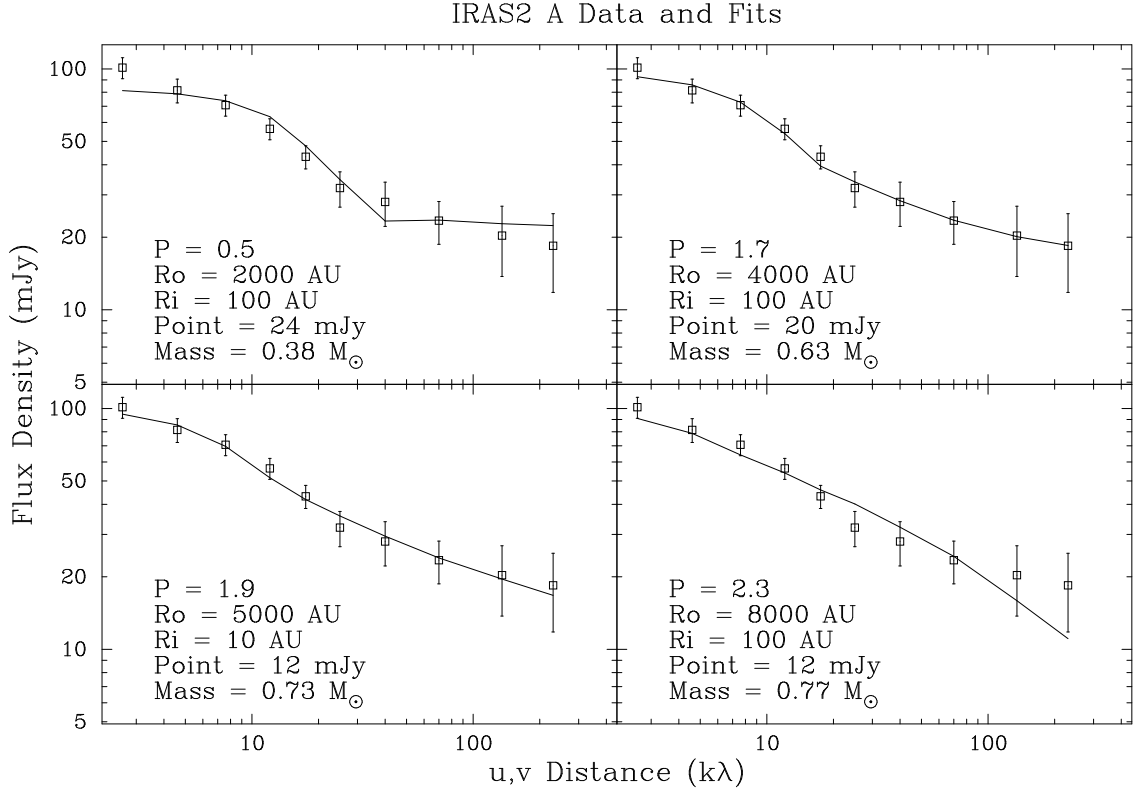}
\vspace{6cm}
\caption{
The observational \uv flux density data binned in annuli around IRAS 2 A and four fits to the
data using the power law envelope
$\rho \propto (\frac{r}{1 AU})^{-p}$ plus a point source.
The parameters for each
fit are listed in the panels. The reduced $\chi^2$ for
all of these fits is less than 1.5.
A wide range in $p$ can be fit by varying the point source flux and
outer radius cutoff to compensate for changing p.
The best fits occur for $p = 1.7 - 1.9$;
$p = 0.5$ and $p = 2.3$ are barely statistically acceptable.
}
\label{iras2afit}
\end{figure}

\begin{figure}
\includegraphics{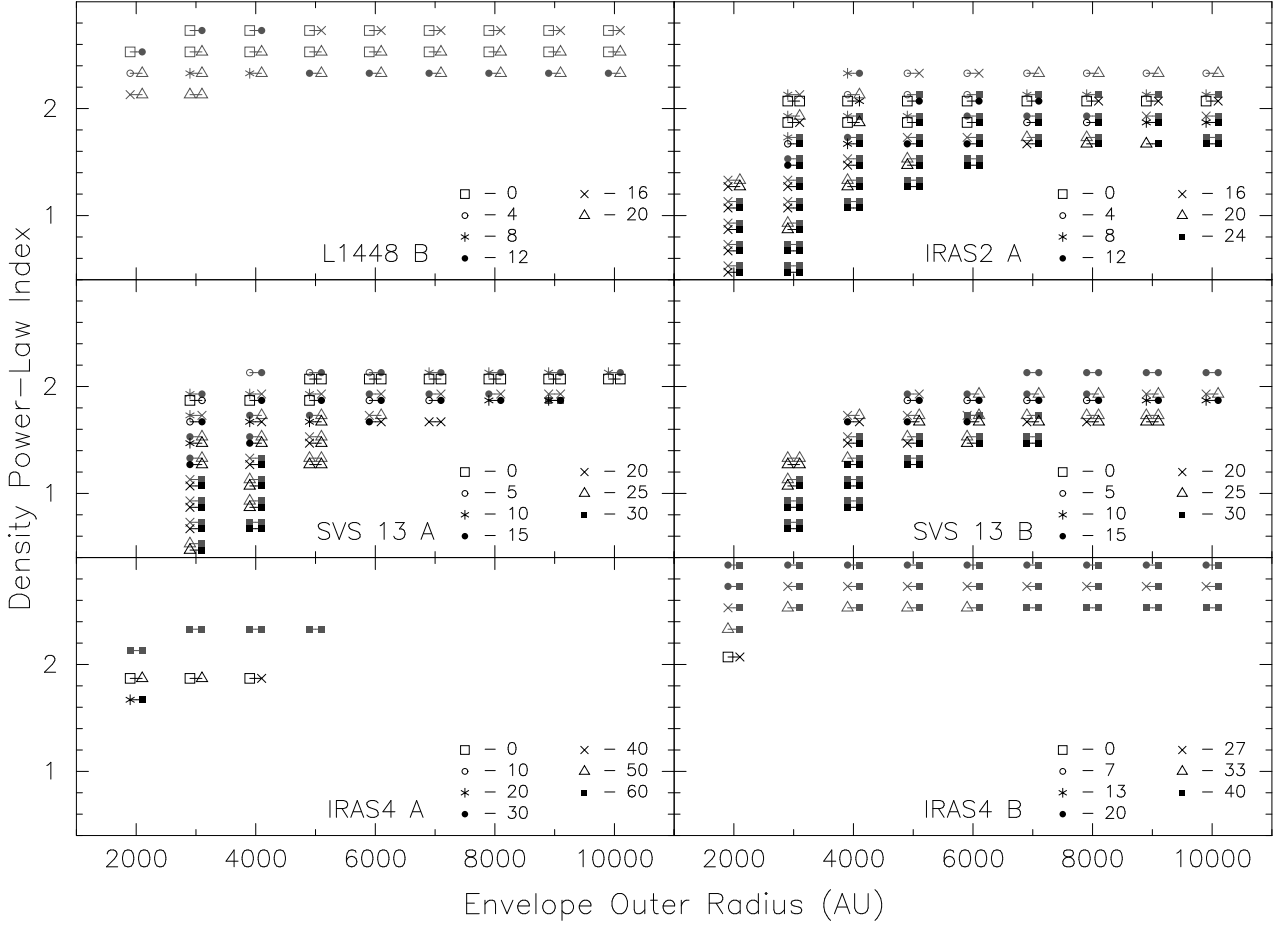}
\vspace{8cm}
\caption{
Fit summary for the six fully modeled sources: L1448 IRS3 B, NGC 1333 IRAS2 A, NGC 1333
SVS 13 A, SVS 13 B, NGC 1333 IRAS 4 A, and IRAS 4 B.
The horizontal axis is the envelope outer radius in AU, and the
vertical axis is the density power-law index.
The black symbols indicate models with an inner radius of 10 AU, while
gray symbols represent models with an inner radius of 100 AU.
For each model with a $>95$\% confidence level in $\chi^2$,
a bar is shown with a symbol at each end that indicates the allowed
range of point sources for that model; a legend to the symbols
is given in the lower right corner of each panel with the flux density in mJy.
}
\label{fitsum}
\end{figure}

We were able to find acceptable
fits for a range of parameter values, but the
data do not always well constrain $p$, even with the highest signal~to~noise
(S/N) sources (Table \ref{fits}).  Fig. \ref{iras2afit}
shows the best fits for NGC 1333 IRAS2~A, a modest S/N example,
for different values of $p$; 
the data can be acceptably fit by $p$ in the range 0.5 to 2.3.
Fig. \ref{fitsum} presents the range of the acceptable
models for the six sources.  
The fits follow general trends, and there are typically a number
of models with good fits within a region of parameter space.
Although most of the envelope models
are fit with density power-law indices
between $p$~=~1.5 and 2.0, some sources prefer steep power-law indices.
In particular, the three sources with the highest S/N only have
acceptable fits for $p > 1.7$. 

In general, but especially in modest S/N cases,
a range of fits are possible as a result of
(1) the outer radius of the envelope 
is typically weakly constrained for models with $p > 1.5$ since
the combination of the density and temperature laws ($p+q$) cause
the total emission to grow slowly with increasing radius,
(2) an increasing point source flux can be compensated by increasing
the inner radius cutoff of the envelope,
and (3) the outer radius and power-law index can
conspire to fit the inner \uv region.
If one or more of these parameters is constrained, the range in
$p$ can become much smaller.
Nonetheless, the higher S/N systems do constrain the power-law in the
envelope, and these data can be used to make model significant statements.

\subsection{Results for $p~=~1.5$ and $p~=~2.0$ }

Now that we have an idea of how constrained the 
$p$ values are by our data, we can explore in more
detail $p~=~1.5$ and $p~=~2.0$ models.
As discussed in \S \ref{intro},
we expect a density structure of
$\rho \propto$~r$^{-2}$ at young ages evolving
from the inside-out at later times to $\rho \propto$~r$^{-3/2}$.
The other parameters are explored over a finer grid:
inner radii ($R_{in}$) of 10, 30, 50 and 100 AU,
outer radius ($R_{out}$) from 1000 AU to
10000 AU, in steps of 1000 AU, and the same step size
for the embedded point source as in the general cases.

\begin{figure}
\includegraphics{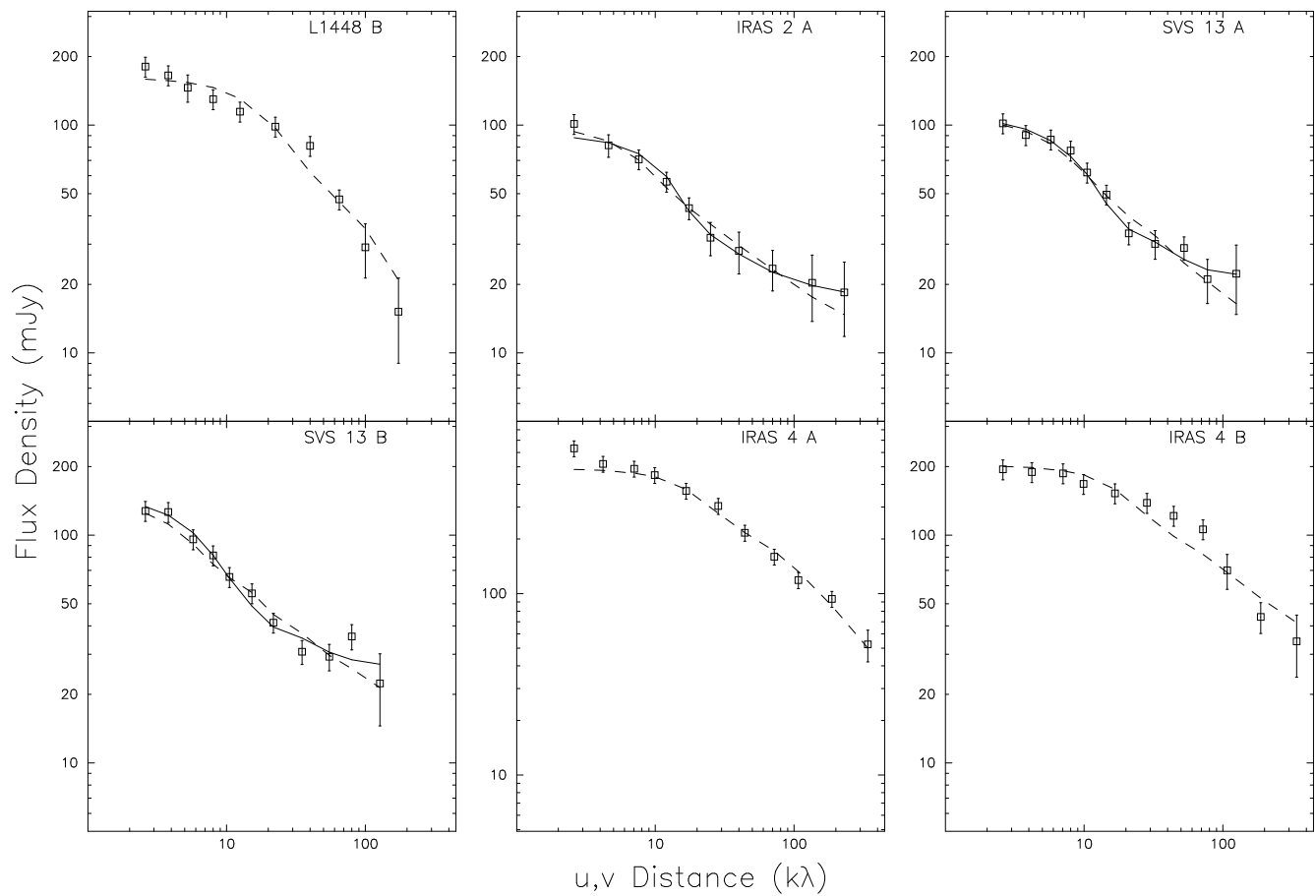}
\vspace{6cm}
\caption{
Examples of best fits for models having power-laws of 
$p~=~1.5$ and $p~=~2.0$ for the six sources that were fully fit. 
The dashed line is the $p~=~2.0$ fit and the solid line is the 
$p~=~1.5$ fit. These fits are representative of a number of fits with
slightly different parameters.
The parameters for the fits displayed are given in Table 2.}
\label{p1.5.2.fits}
\end{figure}

\begin{figure}
\includegraphics{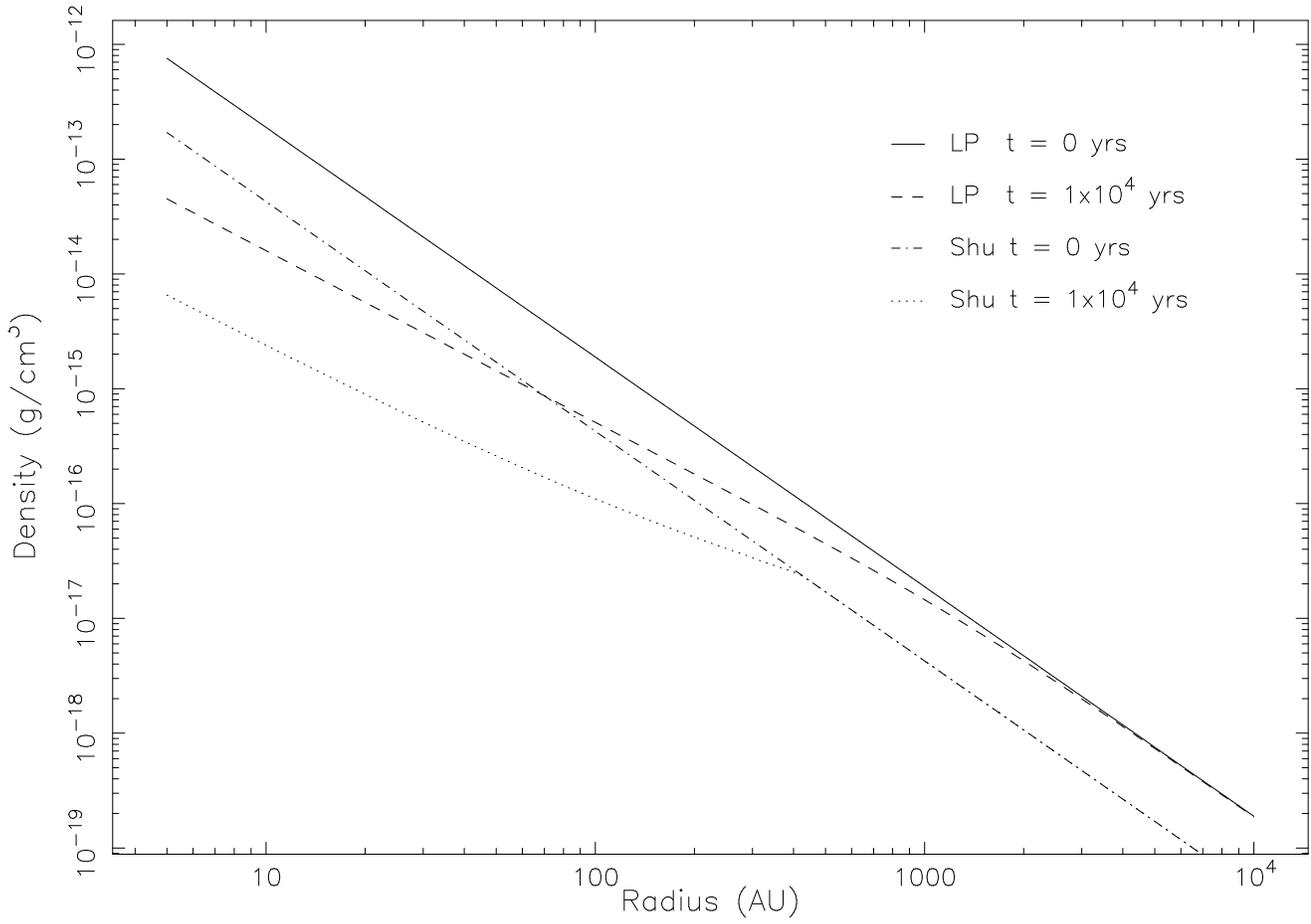}
\vspace{6cm}
\caption{
The Larson-Penston and Shu solutions for the density distribution for 
the Singular Isolated Sphere collapse problem at t~=~0 and 
t~=~1 $\times$ $10^4$ years.
These curves were derived from numerical solution of the differential
equations as discussed in Hunter (1977). The legend in the upper right
identifies the curves.
}
\label{lpshu}
\end{figure}

Fig. \ref{p1.5.2.fits} shows the best fit $p~=~1.5$ and $p~=~2.0$ models
along with the \uv data for the six modeled
sources; the best fit parameters are listed in Table \ref{pfit}.
All six of the sources can be fit with a density power-law index of $p~=~2.0$;
in half of the systems, those with the
highest S/N ratios, $p~=~1.5$ power-law
models can be rejected at the $>95$\% confidence level.
What do these fits tell us about the nature of the systems?
In order to understand the constraints that these higher S/N
systems place on descriptions of the envelope collapse,
it is necessary to examine the Larson-Penston (LP) and Shu solutions 
in more detail.
Fig. \ref{lpshu} shows the density profiles for the two solutions  at
time zero and 10$^4$ years,
assuming a sound speed of 0.2 km/s \citep[e.g.][]{myers83}.
There are two obvious differences: the LP solution has a factor 
of 4.4 higher density 
and the inner envelope region evolves more quickly 
into a $p~=~1.5$ power-law density profile.

In the Shu model, an expansion wave travels from the inner region
outwards--- the ``inside-out'' collapse.
The head of the expansion wave is roughly where the $\rho \sim r^{-2}$
region transitions to the $\rho \sim r^{-3/2}$ region,
the kink near 400 AU radius in Fig. \ref{lpshu}.
The head of the expansion wave is only dependent upon the
time since the formation of the singular core and the 
local sound speed.
Thus, for a sound speed, $a = 0.2$ km/s,
the head of the expansion wave will be located
at 400 and 4000 AU
for ages of $10^4$ and  $10^5$ years, respectively.
Note that in Fig. \ref{lpshu} there
is a significant interface region (100 to 400 AU) in the Shu solution 
where the density slope is shallower than the $p~=~1.5$ power-law, whereas the
LP solution gently increases from $p$~=~1.5 to 2.

In the LP solution, there is no ``inside-out'' expansion wave
since the initial velocity profile is already collapsing,
not at rest as in the Shu solution.
Nonetheless, there is an infall scale for the interface 
between the $\rho \sim r^{-2}$ and $\rho \sim r^{-3/2}$ regions.
Since the LP infall velocity is 3.3 times that in the Shu solution,
the effective spatial scale is 3.3 times as large for the
same sound speed.
Thus, the radius at which one expects a p~=~1.5 is 
$\sim$ 1300 and 13000 AU for ages of $10^4$ and $10^5$ years, respectively.
In addition, the larger mass infall rate creates a higher density envelope
as well as a larger mass protostar earlier in the collapse;
the mass of the envelope at the onset
of the core formation is 4.4 times larger than in the Shu solution.

In order to match the $\rho \sim r^{-2}$
power-law fits, the infall regions in these
models must not dominate the radial density profiles within
a few hundred AU. This implies that
these systems must be young, which is consistent with their classification
as Class 0 sources. The youthfulness of the 
systems is also supported by the dynamical age estimates for the outflows.
For example, L1448 IRS3 has the largest outflow, 14.5$\arcmin$
\citep{bally97}.
For an assumed jet speed of $\sim$ 200 km/s, the minimum age of L1448
IRS3 is $\sim$ 6,200 yrs. The corresponding times for the other systems
are shorter, but it must be stressed that these are {\it minimum} ages as
the outflow region should propagate out at a fraction of the basic
jet speed. 
In addition, all current outflow mechanisms require the
presence of a stellar-size object which is not present until 
after the collapse of the
first hydrostatic core. This collapse occurs when about 0.05 \msun\
has accumulated within 5 AU, effectively extending the
age of these systems with respect to the classic
self-similar solutions \citep[e.g.][]{masunaga98}.

A simple interpretation of these arguments for the infall 
region size would suggest that our fits
to the \uv data require an age of $<$~5000 yrs for the Shu model,
and an age of $<$~1500 yrs for the LP model (assuming $a~=~0.2$ km/s and 
ignoring any differences in absolute density).
At such young ages, the luminosity of the system is also a constraint
since it is determined by the infall rate and the mass of the stellar 
object.
In the above example of L1448 IRS3, which has an estimated luminosity
of 6.8 \lsun, the Shu solution underestimates the luminosity; a 75\% increase 
in the sound speed is needed to match the luminosity for an age of 6,200 years.
Of course, if the sound speed is increased, the expansion wave moves faster and
the radius of the transition 
region increases; this pushes allowed solutions to even earlier times.
The luminosity is not as much of a problem for the LP solution
since its mass accretion is nearly 46.9 times larger at early times. 

\subsection{Results with Fits using the LP and Shu Collapse Solutions}

As the above reasoning suggests, if one wants to compare
the LP and Shu solutions, the age and sound speed must be
included with a complete description of the two solutions.
We have performed modeling of the data using the 
LP and Shu solutions for the density structure of the envelope as a function
of time and sound speed.
Reasonable fits for both models
can be achieved for an age around 1000 years, with $a$~=~0.46~km/s and 0.22 km/s 
for the Shu and LP models, respectively. For example in the source
IRAS~4A, acceptable fits 
were found for ages around 1000 years for the Shu model and 1000~-~2000
years for the LP model. 
These results emphasize that the expected difference in the
extent of the infall region of the two solutions
at young ages is neutralized by the high
sound speed necessary for the Shu solution to fit the physical
densities inferred from the observations.
We stress that we do not propose that the sources
are 10$^3$ yrs old, but show that the strict fit to the data require 
the standard models to be that age.

In our modeling, the LP solution provides an overall better match to
the data, but neither solution is the ideal fit.
The computational simulations of the collapse of isothermal
spheres demonstrate that it is incorrect to think of the envelopes
of young stars as the realization of either collapse model because
the collapse behavior ranges between the LP solution
at early times and the Shu solution at later times
\citep[cf.][]{hunter,foster}.
With that in mind, our modeling of small scale structure in the envelopes of young
stars suggests that 
additional physics is necessary to model the structure of the envelope.
Most likely, inclusion of magnetic fields
\citep[e.g.][]{mous76} and/or turbulence \citep[e.g.][]{gammie} is
important to understand properly the structure in these systems.
Turbulent creation of the initial mass concentration which then evolves
into the core and forming star \citep{eve}
is one example of an intriguing alternative model. 

\subsection{Alternative Assumptions in the Model}

Is there another way in which a $p~=~1.5$ power-law can mimic a
steeper power-law in our fits?
Two of our simplifying assumptions can 
have an impact on the fitted power-law index of the density:
the dust opacity and the geometry.
We adopted a constant dust opacity in our models.
However, dust properties can change with environment
\citep[e.g.][]{gehrz,wein89,henn95}.
There are several grain alterations that may explain an increased dust opacity
in circumstellar regions,
such as chemical evolution (Begemann et al. 1994; van Dishoeck \& Blake 1988), 
formation of dirty ice mantles (Draine 1985; 
Henning, Chini, \& Pfau 1991; Preibisch et al. 1993),
altering of the grain geometry (long ``needle-like'' grains; Wright 1982),
or grain coagulation into fluffy grains (Wright 1987; Jones 1988;
Bazell \& Dwek 1990; Ossenkopf 1991; Stognienko, Henning, \& Ossenkopf 1995).
In any of these scenarios we may expect the dust opacity in the
circumstellar region to become a function of radius: the outer
regions could have grain properties similar to the interstellar medium
while the inner, denser portion of the envelope is likely
to have the most processed, perhaps larger grains due to the 
short timescales for grain alteration.

Since this timescale depends on the density, the 
dust opacity could have a power-law dependence on radius.
In the standard power-law envelope model, the emission is dependent 
on the optical depth, 
$d\tau~=~\kappa_{\nu}~\rho(r)~dl.$
With grain alteration, the optical depth could become the product of 
power-laws in density and dust opacity,
$$d\tau~=~\kappa_{0}\biggl(\frac{\nu}{\nu_{o}}\biggr)^{-\beta}
\biggl(\frac{r}{r_{o}}\biggr)^{-s}\rho_{o}
\biggl(\frac{r}{r_{o}}\biggr)^{-p}~dl,$$
where $s$ is the radial power-law index for optical depth.
Thus, a radial dependency in the dust opacity could be
indistinguishable from that in density and erroneously 
produce steeper density power-law indices in our simple model fits.

\begin{figure}
\includegraphics{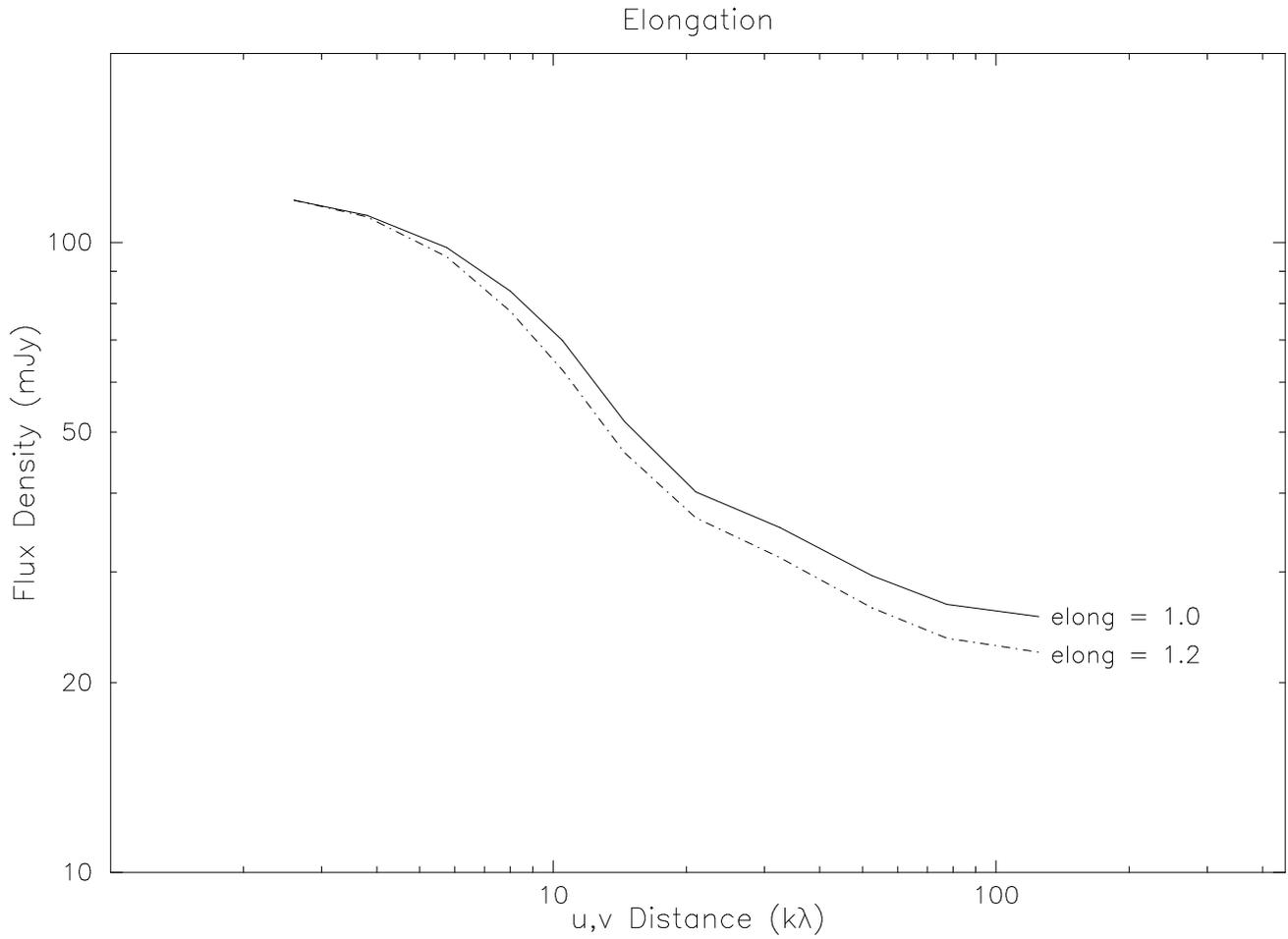}
\vspace{7cm}
\caption{
The effect of elongation of the source on the \uv plane structure. 
The basic model is a power law envelope with $p~=~1.5$, R$_{in}$~=~10~AU,
R$_{out}$~=~4000~AU, and S$_{ps}$~=~20~mJy.
The elongation is specified as the ratio of the major to minor axis 
We increased the major axis relative to the symmetric model.
}
\label{elong}
\end{figure}

Another possible explanation for 
the steeper density profiles is non-spherical or more complicated 
geometries.
All of the embedded systems are known to be multiple systems
or driving large molecular outflows
that are evacuating material out of the envelope,
both of which may significantly affect the envelope
structure.
In order to explore the effect of geometry on our models,
we stretched one of our envelope models such that the elongation
ratio of the semi-major to semi-minor axis is 1.2.
Fig. \ref{elong} shows that in this simple model,
an elongation can make the density
power-law index appear steeper than expected.
Note however that the observed emission from the sources modeled
here are roughly circular symmetric.

\section{Conclusions}

We have presented the first modeling of circumstellar
envelope dust emission that is uniformly sensitive to
core structure from sub-arcsecond out to 50 arcseconds. 
Other groups have done modeling of the
density structure in the outer envelope 
(c.f. Chandler \& Richer 2000; Shirley et al. 2000),
but probing the inner envelope, where the collapse 
solutions differ, requires high spatial resolution.
Even with our high spatial resolution, 
only high signal-to-noise data can provide significant 
constraints to the standard collapse models.
With the models presented in this paper, we place some of the first
constraints on the emission contributions from the envelope, inner
envelope, and disk in the youngest protostellar sources 
in the nearest star forming regions.
Morphologically, the \uv data of all the sources typically show
extended objects that either appear to follow a power-law down to small
scales (large \uv) or level out at some constant residual value, 
presumably a disk.

When modeling the inner envelope with high resolution, the standard
power-law temperature model is not valid, and a self-consistent,
luminosity preserving technique is required.
Using a self-consistent temperature model and a dust structure model
consisting of a simple power-law density, inner and outer radius cutoffs,
and an embedded point source to emulate an embedded
circumstellar disk, we have fit the \uv data from our $\lambda$~=~2.7~mm
sub-arcsecond survey (LMW), placing constraints upon the 
conditions in the early stages of star formation.

Specifically, we found that the density structure
in  L1448 IRS3 B, NGC 1333 IRAS 4 A, and NGC 1333 IRAS 4 B
are best-fit with a power-law index of p~=~2.0.
However models of star formation (Larson-Penston and Shu solutions
to the isothermal sphere) predict that
a $\rho ~\propto~ r^{-3/2}$ density power-law should be established
quickly in the collapse process.
So neither of the two self-similar collapse solutions
exactly fit our data, although the LP solution provides the overall
better fit with more likelihood and allowing the observed luminosity.  
The strict Shu inside-out collapse solution 
fit to the data requires very young, unphysical
ages ($<$ 1000~yrs) and very high sound speeds that
in some cases are still not enough to match the luminosity.
The LP solution easily fits the luminosity
constraints, but the LP solution evolves too quickly into a 
$\rho ~=~ r^{-3/2}$ density power-law.
The strict LP solution fit to the data also requires
very young ages 1000 to 2000 yrs that are non-physical: all
of the modeled systems have central stellar objects and outflows.
Age estimates based on the apparent outflow extent
suggest typical ages for these systems of one to a few times 10$^4$ years.

It is clear that there is a fundamental problem with
the time scale in these simple self-similar models; either the
models are not accurately including all the physics important to evolution at
early times or in a real system the t~=~0 point is not well determined. 
Incorporation of magnetic fields and turbulence, on all scales, is
likely to be important new pieces.
Such an improvement of models will become even more critical
in the future with the advent of higher resolution and higher sensitivity
submillimeter and millimeter observations become available from
the SMA, CARMA, and ALMA.

One of the primary strengths of this study is the 
ability of the interferometer to separate 
large scale emission from compact emission, allowing us
to probe for a circumstellar disk 
component embedded within a circumstellar envelope.  
Our data solidly show that most of the emission, a minimum of
85\% up to 100\% arises from the circumstellar envelope.
But, point source flux values in the range of 0 to 40 mJy 
are found for the best fits (Table \ref{fits}).
We can make a mass estimate for an embedded disk using the
circumstellar disk of HL Tauri as a standard.
A HL Tauri type disk (disk mass $\sim$~0.05~\msun; LMW)
has a flux of $\sim$~100~mJy at the distance of Taurus (140 pc).
If placed at the distance of Perseus (350 pc), the flux of HL Tauri
would be 16 mJy.
So the range of acceptable point source fluxes would represent 
circumstellar disk masses of 0 to 0.12~\msun.
This is a small fraction of the circumstellar envelope mass,
typically $\sim$~1~\msun.
In other words, since the disks are not over-massive 
compared to the Class I/II systems, the larger quantity of
infalling material in the Class 0 stage must
process through the disk and onto the protostar quickly.

\acknowledgments
We especially thank Mark Wolfire for all of his help and advice on
his self-consistent radiative transfer code without which this paper
could never have been realized.
We thank Pedro Safier for discussions on cloud collapse and
Eve Ostriker, Steve Lubow, Chris Hunter, and Fred Adams for useful discussions.
LWL would also like to thank the Infrared Group at MPE and its Director Reinhard Genzel in particular.
This work was supported by NSF Grants NSF-FD93-20238 and AST-9314847. 
LWL and LGM acknowledge support from NASA grant NAGW-3066.

\clearpage

\clearpage

\clearpage

\begin{deluxetable}{llccccc}
\tablewidth{0pt}
\tablecaption{Summary of Fits for Variable Density Power Law}
\tablehead{
  \colhead{} & \colhead{} &  \multicolumn{5}{c}{Most Likely Parameters}\\
  \colhead{Source}       &
  \colhead{$p$-index}      &
  \colhead{$p$-index}      &
  \colhead{$M_{env}$}    &
  \colhead{R$_{outer}$} &
  \colhead{$S_{ps}$}     &
  \colhead{$M_{disk}$ \tablenotemark{\dagger}}   \\
  \colhead{ }            &
  \colhead{(range)}      &
  \colhead{}             &
  \colhead{(\msun)}      &
  \colhead{(AU)}         &
  \colhead{mJy}          &
  \colhead{(\msun)}
}
\startdata
L1448 IRS3 B      &  2.1 - 2.7   & 2.5  &  1.63 & 8000 & 12 & 0.03 \\
NGC 1333 IRAS2 A  &  0.5 - 2.3   & 1.7  &  0.63 & 4000 & 20 & 0.06 \\
SVS 13 A          &  0.5 - 2.1   & 0.9  &  0.60 & 3000 & 25 & 0.08 \\
SVS 13 B          &  0.7 - 2.1   & 1.7  &  1.06 & 5000 & 20 & 0.06 \\
NGC 1333 IRAS4 A  &  1.7 - 2.3   & 1.9  &  3.60 & 2000 & 0  & 0.00 \\
NGC 1333 IRAS4 B  &  2.1 - 2.9   & 2.9  &  1.57 & 7000 & 40 & 0.12
\tablenotetext{\dagger}{For $M_{disk}$ mass, assumed HL Tauri-like
circumstellar disk -- at a distance of 140 pc,
$M_{disk}$ = 0.05 \msun with 100 mJy emission at $\lambda$~=~2.7~mm.}
\enddata
\label{fits}
\end{deluxetable}
\clearpage

\clearpage

\clearpage

\clearpage

\begin{deluxetable}{lcccccc}
\tablewidth{0pt}
\tablecaption{$p$ = 1.5 and 2.0 Best Fits}
\tablehead{
  \colhead{Source}       &
  \colhead{p-index}      &
  \colhead{M$_{env}$}    &
  \colhead{R$_{in}$}     &
  \colhead{R$_{out}$}    &
  \colhead{S$_{ps}$}     &
  \colhead{$\chi _{red} ^{2}$}    
}
\startdata
L1448 IRS3 B   &  1.5  & \nodata & \nodata & \nodata & \nodata & \nodata \\
               &  2.0  & 0.97    & 30      & 2000    & 0       & 1.4     \\
IRAS 2 A       &  1.5  & 0.52    & 100     & 3000    & 20      & 0.3     \\
               &  2.0  & 0.68    & 100     & 5000    & 16      & 0.3     \\
SVS 13 A       &  1.5  & 0.78    & 10      & 4000    & 20      & 0.3     \\
               &  2.0  & 0.91    & 10      & 5000    & 0       & 0.7     \\
SVS 13 B       &  1.5  & 1.13    & 10      & 5000    & 25      & 0.7     \\
               &  2.0  & 1.26    & 10      & 7000    & 5       & 1.1     \\
IRAS 4 A       &  1.5  & \nodata & \nodata & \nodata & \nodata & \nodata \\
               &  2.0  & 3.37    & 30      & 2000    & 10      & 0.4     \\
IRAS 4 B       &  1.5  & \nodata & \nodata & \nodata & \nodata & \nodata \\
               &  2.0  & 1.65    & 30      & 2000    & 33      & 1.4     
\enddata
\label{pfit}
\end{deluxetable}
\clearpage

\clearpage

\clearpage

\end{document}